\documentclass[debug]{rmaa}

\usepackage{paralist}
\usepackage{{longtable}}

\usepackage{psfrag,color}

\usepackage[latin1]{inputenc}

\newcommand{\msec}[2]{$#1\mbox{$''\mskip-7.6mu.\,$}#2$}

\title{Radio emission variability and proper motions of WR 112} 

\author{
  J. O. Yam\altaffilmark{1},
  S. A. Dzib\altaffilmark{2},
  L. F. Rodr\'{\i}guez\altaffilmark{3,4} and
  V. Rodr\'{\i}guez-G\'omez\altaffilmark{5}
  }

\altaffiltext{1}{Departamento de Ciencias, Universidad de Quintana Roo, M\'exico.}

\altaffiltext{2}{Max-Planck-Institut f\"ur Radioastronomie, Bonn, Germany.}
\altaffiltext{3}{Centro de Radioastronom\'{\i}a y Astrof\'{\i}sica,
  UNAM, M\'exico.}
\altaffiltext{4}{Astronomy Department, Faculty of Science, King Abdulaziz University, Jeddah, Saudi Arabia}
\altaffiltext{5}{Harvard-Smithsonian Center for Astrophysics, USA.}
\shortauthor{Yam et al.}
\shorttitle{WR112}

\fulladdresses{
\item V. Rodr\'iguez-G\'omez: Harvard-Smithsonian Center for Astrophysics, 60 Garden Street, Cambridge, MA 02138, USA.
\item S. Dzib: Max-Planck-Institut f\"ur Radioastronomie, Auf dem H\"ugel 69, 53121 Bonn, Germany (sdzib@mpifr-bonn.mpg.de).
\item L. F. Rodr\'iguez: Centro de Radioastronom\'\i{}a y
  Astrof\'\i{}sica, Universidad Nacional Aut\'onoma de M\'exico,
  Apartado Postal 3--72, 58090 Morelia, Michoac\'an, M\'exico and Astronomy Department, Faculty of Science, King Abdulaziz University, P.O. Box 80203, Jeddah 21589, Saudi Arabia (l.rodriguez@crya.unam.mx).
\item Omar Yam: Universidad de Quintana Roo, Boulevard Bah\'ia S/N, C. P. 77019, Chetumal, Quintana Roo, M\'exico
  (oyam@uqroo.edu.mx).}

\listofauthors{Author1, Author2, \& Author3}
\indexauthor{Author1, W. J.}
\indexauthor{Collaborator, A.}
\indexauthor{Author, L.}

\resumen{Analizamos 64 observaciones en radio a la frecuencia de 8.4 GHz de la estrella
Wolf-Rayet WR~112, tomadas de los archivos del {\it Very Large Array}.
Las observaciones cubren una l\'{\i}nea de base temporal de 13 a\~nos,
de junio del 2000 a julio de 2013. La estructura
de WR~112 es consistente con ser una fuente puntual en todas las \'epocas y su densidad de 
flujo var\'\i a entre 0.6 mJy y 2.1 mJy. Intentamos buscar periodicidades
en estas variaciones sin \'exito.
Tambi\'en buscamos emisi\'on extendida asociada con la nebulosa infrarroja que rodea a WR~112,
poniendo l\'\i mites superiores de
50 $\mu$Jy. Finalmente, usamos
las observaciones con resoluci\'on m\'as alta para medir los movimientos
propios de WR~112, obteniendo
$\mu_\alpha\cos \delta = -2.6 \pm 1.1 \mbox{~mas~yr$^{-1}$}$, y
$\mu_\delta  = -5.4 \pm 1.4 \mbox{~mas~yr$^{-1}$}$.  Estos movimientos
propios son menores que los reportados previamente, pero contin\'uan sugiriendo movimientos
peculiares significativos.}

\abstract{We analyzed 64 radio observations at the frequency of 8.4 GHz of the Wolf-Rayet
star WR~112, taken from the {\it Very Large Array} archive. These observations
cover a time baseline of 13 years, from June 2000 to July 2013. 
The radio structure of WR~112 is consistent with it being a point
source in all the epochs and with its flux density varying from 0.6 mJy
to 2.1 mJy. We tried to search for periodicities
in these variations but our results were not conclusive. We also looked for
extended emission from the infrared nebula that surrounds WR~112, settimg upper limits of
50 $\mu$Jy. Finally, we used the highest angular resolution
images to measure the proper motions of WR~112, obtaining
$\mu_\alpha\cos \delta = -2.6 \pm 1.1 \mbox{~mas~yr$^{-1}$}$, and
$\mu_\delta  = -5.4 \pm 1.4 \mbox{~mas~yr$^{-1}$}$. These proper motions
are smaller than those previously reported, but still suggest significant peculiar
motions for WR 112.}

\addkeyword{circumstellar matter}
\addkeyword{astrometry}
\addkeyword{stars: Wolf-Rayet}
\addkeyword{stars: individual: (WR 112)}

\begin{document}
% Typeset article header
\maketitle

\section{INTRODUCTION}
\label{sec:intro}

A small fraction of the Wolf-Rayet (WR) stars are known to be dust-producing
sources. How the dust is formed in these hostile environments 
and whether or not is this dust production episodic or continuous are still
not understood issues.  Among this dust-producing WR stars we have WR~112, which belongs 
to the carbon subclass and has been classified as WC9 
\citep{1983PASP...95..440M}. Marchenko et al.~(2002) suggested that 
it is a binary system with a period of 25 years, with the
companion being an OB star. However, 
there has not been a direct detection of this proposed companion.
WR~112 is at a distance of 4.15 kpc~\citep{2001VisieR}.
The dust production rate for WR~112 is
estimated to be $\sim 6.1\times 10^{-7}M_\odot$ 
yr$^{-1}$~\citep{2002ApJ...565L..59M}. The dust exhibits a pinwheel-like
morphology around WR~112, at arcsecond scales (Marchenko et al.~2002),
that could extend down to sub-arcsecond scales (i.e., Ragland \& Richichi
1999). WR~112 also shows variable
radio emission \citep{2002ApJ...566..399M} and significant proper motions 
\citep{2009RMxAA..45....3D}. Whether this radio emission is periodic or not
and how it is related with its proper motions, the periods of the
possible companions and the periods of dust formations are points that remain unknown.

WR~48a is a carbon-rich WR star, also known as a long-period dust
maker and a colliding-wind binary, and thus similar to WR~112.
Multi--epoch studies of WR~48a with infrared photometry by 
\citet{2012MNRAS.420.2526W} imply 
a period for dust emission of $\sim$ 32 years and that there are also some irregular 
\emph{mini}-periods of a few years.
Determination of the period of the radio variations and the proper motions
of WR~112 could give insights into the orbital elements
of the companion  and how it is related to the dust production.
In the present work we will use radio observations taken over a period of 13 years to study the enigmatic
WR~112 system.

\section{OBSERVATIONS AND DATA CALIBRATION}
\label{sec:errors}

The radio observations of WR~112 were taken from the archives of the Very
Large Array (VLA) of the NRAO\footnote{The National Radio Astronomy
Observatory is operated by Associated Universities Inc. under cooperative
agreement with the National Science Foundation.}. We found a total
of 64 observations in the VLA archive at a frequency of 8.4 GHz 
($\lambda\,=\,$3.6 cm)\footnote{We found a few other observations at other
frequencies, but they were not useful for the purposes of the present work,
thus we did not take them into account.}. These observations cover the period
from June 2000 to July 2013. For some epochs two phase calibrators, 
J1820-254 and J1832-105, were used. J1820-254 is at an angular distance
of $6\rlap.^\circ57$ from WR~112, and showed a flux density in the range of 0.63 to 1.20 Jy. Similarly,
J1832-105, is at an angular distance of $9\rlap.^\circ22$ from WR~112, and showed 
a flux density in the range of 1.29 to 1.54 Jy.
To obtain the best gain transfers it is recommended to choose a phase
calibrator close to the target, with a large flux density and 
pointlike morphology. Furthermore, to obtain accurate absolute astrometry
the same phase calibrator must be used in all epochs.  Because J1820-254 has
a better determined position and it is closer to WR 112 than
J1832-105, our analysis is restricted only to the observations that are phase
calibrated with J1820-254. Fortunately, this calibrator is present in all
the epochs.  

The data were edited, calibrated and imaged in the standard fashion using
the Common Astronomy Software Applications package (CASA). After the initial 
calibration, the visibilities were imaged with a pixel size of a fifth the
size of the resulting synthesized beam at each epoch. The weighting scheme
used was intermediate between natural and uniform (WEIGHTING='briggs' with
ROBUST=0.0 in CASA). The rms noise level of the images is typically 
$\sim50$ $\mu$Jy beam$^{-1}$, but it can be slightly different for some
epochs.

\section{Results and Discussions}
\label{sec:EPS}

\subsection{Radio Emission Variation}

WR~112 was detected in all the epochs with a point-like structure.
Its flux density at each epoch was measured
using the task IMFIT in CASA, and the values are given in Table~\ref{tab1}.
The flux densities of WR 112 show variations at levels from
0.58 mJy up to 2.08 mJy (see Table~\ref{tab1}). To determine
whether these variations are periodic or not, we proceeded as follows.

According to \citet{2002ApJ...566..399M}, WR 112 was 
in a very high state of activity reaching up to 4.4 mJy at 8.4 GHz
during 1999 September until after 2000 February. Therefore, 
the  observation of June 2000, corresponding to our maximum measured
flux density of 2.08 mJy, could be a remnant of the activity that
occurred during that period and was not included in the analysis. 

\begin{figure*} 
  \centerline{\includegraphics[width=.8\textwidth,angle=0]{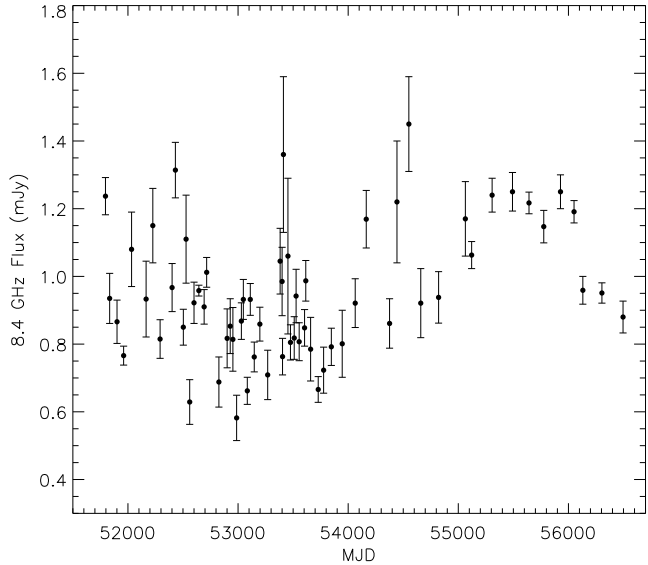}} 
  \caption{Flux densities of WR 112 at 8.4 GHz as a function of MJD.} 
 \label{fig:period} 
\end{figure*}

The flux densities of the remaining observations (plotted in Figure \ref{fig:period}) were
used to look for periodicities in the data using
the Lomb-Scargle method (Scargle 1982; Lomb 1976; Press et al. 1992). This method
estimates a frequency spectrum
for an incomplete or unevenly sampled time series. For this, the method uses least squares fits
of sinusoidal functions over a determined period range. In our case, we used a range from
10 days to 13 years.
We obtained two possible periods, one of 11.7 years and the other of 18.8 days. However, the false alarm probabilities
are 22\% and 23\%, respectively, and thus the possible periodicities are marginal. In conclusion, we did not find any
reliable periodicity in the flux density variations in WR 112.

\subsection{Extended Nebula}

One of the most important features of WR~112 is that it is a dust producer
star and has a circumstellar dust shell detected at 2.2 $\mu$m 
\citep{1999MNRAS.302L..13R}. Due to the morphology of its envelope --a
rotating spiral dust shell extending up to 6\arcsec-- WR 112 has been
classified as a pinwheel nebula. The origin of this structure was probably formed in the
wind-wind collision region of the binary system\citep{2002ApJ...565L..59M}. 
The pinwheel nebula structure could also be present at sizes down to
$\sim20$ mas (Monnier et al.~2007).

In order to know more about the pinwheel structure of WR~112, sensitive 
observations at several wavelengths are needed. The intensity of
dust emission rises rapidly with frequency and it is expected to be more important 
at higher frequencies than 8.4~GHz. However, high sensitivity observations
at higher frequencies have not been carried out, to our knowledge, with the VLA.
Thus, we attempted to detect possible ionized material related to the
pinwheel-nebula at the 8.4 GHz frequency of the observations analyzed.

\begin{figure*}[!h] 
  \centerline{\includegraphics[width=1.0\textwidth,angle=0,clip=true]{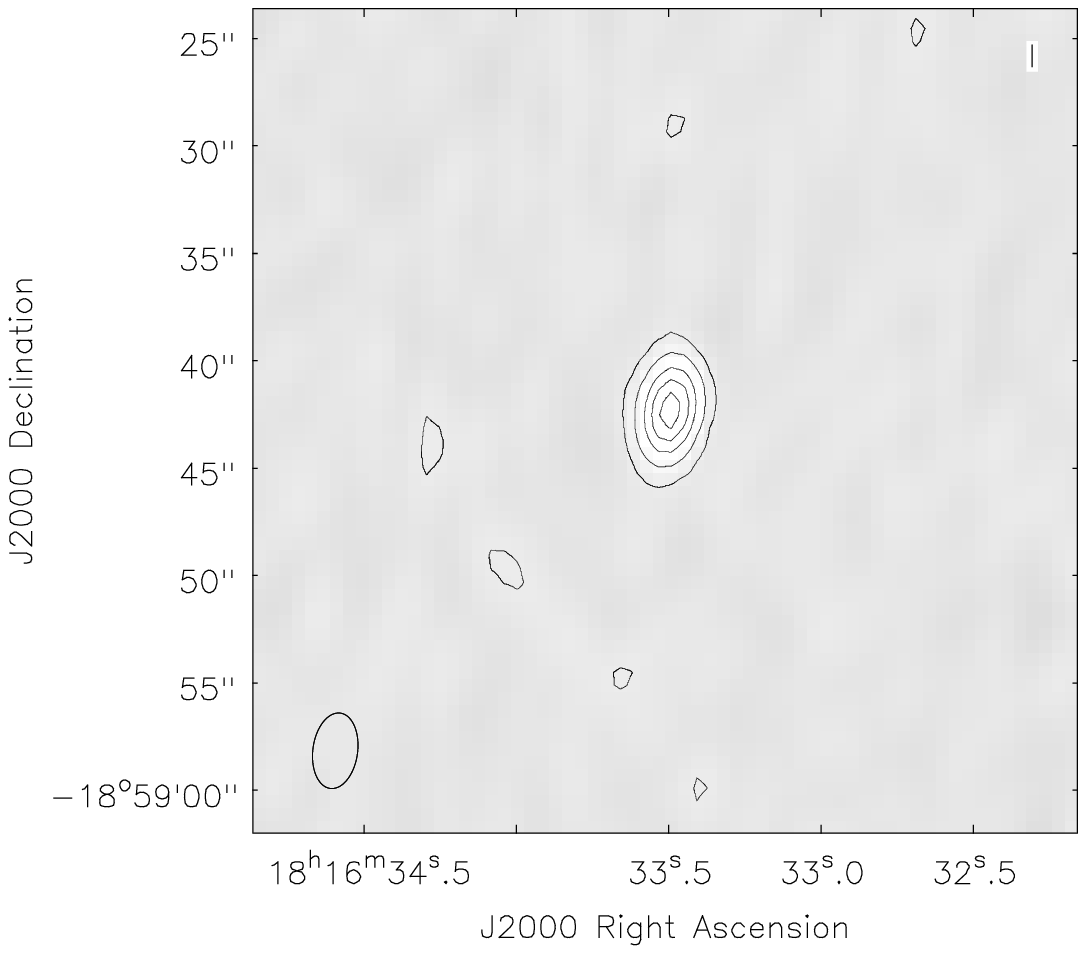}} 
  \caption{Multi--epoch image of WR 112 at 8.4 GHz combining VLA C and D configurations.
The measured flux density is 0.78$\pm$0.02 mJy. Contours levels correspond
to -3, 3, 10, 20, 30 and 40 times 16 $\mu$Jy beam$^{-1}$, the noise level of
the image. The half power contour of the synthesized
($3\rlap{''}.\,5 \times 2\rlap{''}.\,1; P.A. = -8^\circ$)
beam is shown in the bottom left corner.}
 \label{fig2} 
\end{figure*}

This structure is not present in any of our final single--epoch images. 
One way to gain more sensitivity is by combining the visibilities
of the single epochs and produce a multi--epoch image. We notice, however,
that the highest resolution could be affected by movements of the star
and its more immediately surrounding material, see also next section.  
Then, we use the observations of the more compact VLA configurations, C and D.
We concatenated all the observations corresponding to these
configurations and WR~112 still looks as a point source with a
$3\rlap{''}.\,5 \times 2\rlap{''}.\,1$
resolution (see Fig.~\ref{fig2}). 
No extended structures are detected at levels above 50 $\mu$Jy. 
Sensitive images at higher frequencies could detect the dust emission
of the pinwheel nebula in WR~112.

%Para la Configuración D;

%rms=1.8116e-5 Jy/beam
%Beam=3.61" X 2.03"
%Image size=3.65"(+/- 0.29) X 1.95"(+/-0.51)
%peak=0.892+/- 0.018 mJy/beam
%flux=0.867+/-0.017 mJy

%para a configuración D,

%rms=2.1521e-5 Jy/beam
%Beam=3.54" X 2.08"
%Image size=3.55"(+/- 0.34) X 2.06"(+/-0.58)
%peak=0.788+/- 0.021 mJy/beam
%flux=0.781+/-0.021 mJy

\subsection{Proper Motions} 

For the determination of the proper motion we used the data sets
of the epochs with the VLA in its A configuration, the most extended.
We also used the July 2012 (2012.62) epoch, that was obtained in the
B configuration for two reasons. First, it was observed with the new
Karl G. Jansky VLA system, and the signal to noise level reached for
this epoch was better than any of the previous B configuration observations,
and thus we can determine a good position. Second, it helps to have
a longer time baseline, that improves the astrometric measurements.

With the positions presented in Table~\ref{tab2}, we determine the
proper motions of WR 112 to be\footnote{We do not include epochs 2003.79
and 2007.76 in our proper motion analysis because their positions do not
follow the trend presented by the others epochs, with a difference of up to 
$\sim$ \msec{0}{15}.}:

\begin{eqnarray} 
\mu_\alpha\cos \delta  & = &  -2.6 \pm 1.1 \mbox{~mas~yr$^{-1}$}\nonumber\\ 
\mu_\delta  & = &  -5.4 \pm 1.4 \mbox{~mas~yr$^{-1}$.} \nonumber 
\end{eqnarray}

\begin{figure*}[!ht] 
  \centerline{\includegraphics[width=\textwidth,angle=0,clip=true]{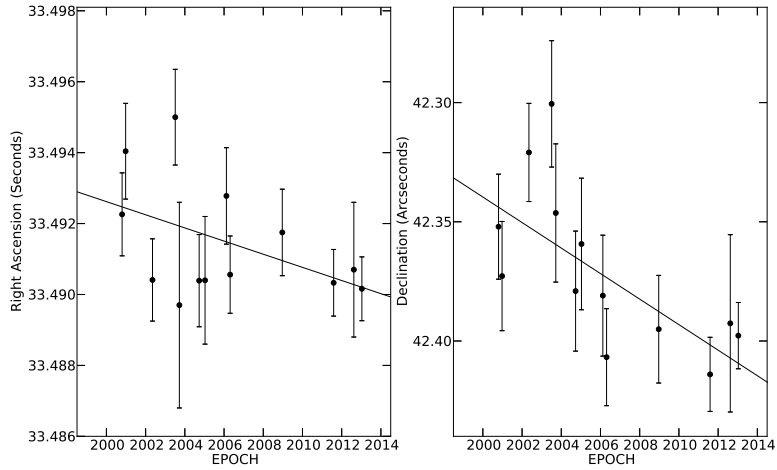}} 
  \caption{Positions of WR~112 as a function of time. The right ascension is $18^h 16^m$ and the declination is -18\arcdeg 58\arcmin . The solid lines are the least squares best fits to the positions.}
%The red points shows the position of the epochs that were not used in the fit. Note that in the
%2003.79 epoch the declination position is too far away from the other positions that is not included
%in the plot.} 
 \label{fig:pm} 
\end{figure*}

Systematic contributions of \msec{0}{010} and \msec{0}{009}
in the $\alpha$ and $\delta$ positions, respectively, 
were added in quadrature to the positional errors obtained from 
a Gaussian fit (task IMFIT in CASA), in order to obtain a $\chi^2$ of 1. 
The positions as a function of the epoch are presented in Figure \ref{fig:pm}. 

The measured proper motions are significantly smaller than those 
found by Dzib \& Rodr\'{\i}guez~(2009): $\mu_\alpha\cos \delta = -11.2 \pm 3.1 
~mas~yr^{-1}$, $\mu_\delta = -13.5 \pm 5.5   
~mas~yr^{-1}$. Dzib \& Rodr\'{\i}guez~(2009) noted that the
movements of WR~112 seemed to be affected by an abrupt change
between two epochs. Inspecting Figure \ref{fig:pm},
we can note that the positions present some dispersion and
that abrupt changes in position are also present. These changes
are more prominent in the declination direction, in agreement
with those noticed by Dzib \& Rodr\'{\i}guez~(2009).
As we have used more epochs than Dzib \& Rodr\'{\i}guez~(2009),
the dispersion is smoothed and allows us to measure more reliable
proper motions. The noise of the new proper motions are 2 to 3 times
smaller than those of Dzib \& Rodr\'{\i}guez~(2009). 

In galactic coordinates the new
proper motions are
$\mu_l~ cos~ b = -6.0 \pm 1.7 
~mas~yr^{-1}$, $\mu_b = -0.3 \pm 0.3 
~mas~yr^{-1}$. At the position of WR~112 the expected proper motions for a
source that is stationary with respect to its local standard of rest
are $\mu_l~ cos~ b = -1.1
~mas~yr^{-1}$, $\mu_b = -0.3
~mas~yr^{-1}$ (Dzib \& Rodr\'{\i}guez 2009). 
Then, WR~112 has residual proper motions of $-4.9 \pm 1.7 
~mas~yr^{-1}$ in galactic longitude and of
$0.0 \pm 0.3 ~mas~yr^{-1}$ in galactic latitude. 
These results are consistent with no peculiar motions in
galactic latitude and a peculiar motion of -100$\pm$34 km s$^{-1}$  in galactic longitude.
Even when the measurement is significant only at the 3-$\sigma$ level, it suggests significant
peculiar motions for WR~112. 
Moffat et al. (1998) give as the criterion for a runaway star to exceed
a velocity of
42 km s$^{-1}$ with respect to its local standard of rest.
Then, WR~112 could be a runaway star and more accurate proper motions
determinations are required to test this possibility.

% Understanding the origin of the position dispersion is
% an interesting subject. Dzib \& Rodr\'{\i}guez~(2009) discussed
% that the origin of the abrupt changes could be due to the
% presence of the pinwheel nebula. The material of this pinwheel nebula
% could interact with the surrounding material and produce
% radio emission, and could also explain the smaller changes in
% positions. However, if this shift in positions are due to the
% presence  of new shock zones, it is not clear how the total
% radio flux does not show large variations. Sensitive, higher angular
% resolution observation could help to solve this problem.

\begin{center}
\begin{longtable}{lrrrcc}
\caption{VLA Observations of WR112 at 8.4 GH\lowercase{z}.} \label{tab1} \\
    \toprule
  Epoch & \multicolumn{1}{c}{MJD} & \multicolumn{1}{c}{Config.} & \multicolumn{1}{c}{Project} & \multicolumn{1}{c}{Flux} \\
dd/mm/yyyy & & & & (mJy)\\
    \midrule
\endfirsthead

\multicolumn{6}{c}%
{{\bfseries \tablename\ \thetable{} -- continued from previous page}} \\
     \toprule
  Epoch & \multicolumn{1}{c}{MJD} & \multicolumn{1}{c}{Config.} & \multicolumn{1}{c}{Project} & \multicolumn{1}{c}{Flux}\\
dd/mm/yyyy & & & & (mJy)\\
    \midrule
\endhead

\midrule \multicolumn{6}{r}{{Continued on next page}} \\ \midrule
\endfoot

\endlastfoot
20/06/2000 &  51715 &  CD &    AM661 & $ 2.08\pm0.04 $   \\
09/09/2000 &  51796 &   D &    AM661 & $ 1.24\pm0.06 $   \\
17/10/2000 &  51834 &   A &    AM661 & $ 0.94\pm0.07 $   \\
23/12/2000 &  51901 &   A &    AM661 & $ 0.87\pm0.06 $   \\
21/02/2001 &  51961 &  AB &    AM661 & $ 0.77\pm0.03 $   \\
04/05/2001 &  52033 &   B &    AM661 & $ 1.08\pm0.11 $   \\
13/09/2001 &  52165 &   C &    AM687 & $ 0.93\pm0.11 $   \\
13/11/2001 &  52226 &   D &    AM687 & $ 1.15\pm0.11 $   \\
18/01/2002 &  52292 &   A &    AM687 & $ 0.82\pm0.06 $   \\
07/05/2002 &  52401 &   A &    AM727 & $ 0.97\pm0.07 $   \\
06/06/2002 &  52431 &  AB &    AM727 & $ 1.31\pm0.08 $   \\
17/08/2002 &  52503 &   B &    AM727 & $ 0.85\pm0.05 $   \\
11/09/2002 &  52528 &   B &    AM727 & $ 1.11\pm0.13 $   \\
14/10/2002 &  52561 &   C &    AM727 & $ 0.63\pm0.07 $   \\
22/11/2002 &  52600 &   C &    AM727 & $ 0.92\pm0.06 $   \\
05/01/2003 &  52644 &   C &    AM727 & $ 0.96\pm0.02 $   \\
22/02/2003 &  52692 &   D &    AM727 & $ 0.91\pm0.05 $   \\
17/03/2003 &  52715 &   D &    AM727 & $ 1.01\pm0.04 $   \\
06/07/2003 &  52826 &   A &    AM766 & $ 0.69\pm0.07 $   \\
19/09/2003 &  52901 &   A &    AM766 & $ 0.82\pm0.09 $   \\
17/10/2003 &  52929 &  AB &    AM766 & $ 0.85\pm0.08 $   \\
10/11/2003 &  52953 &   B &    AM766 & $ 0.81\pm0.09 $   \\
14/12/2003 &  52987 &   B &    AM766 & $ 0.58\pm0.07 $   \\
26/01/2004 &  53030 &  BC &    AM766 & $ 0.87\pm0.05 $   \\
13/02/2004 &  53048 &  BC &    AM766 & $ 0.93\pm0.06 $   \\
20/03/2004 &  53084 &   C &    AM766 & $ 0.66\pm0.04 $   \\
16/04/2004 &  53111 &   C &    AM766 & $ 0.93\pm0.05 $   \\
22/05/2004 &  53147 &   C &    AM793 & $ 0.76\pm0.04 $   \\
12/07/2004 &  53198 &   D &    AM727 & $ 0.86\pm0.05 $   \\
21/09/2004 &  53269 &   A &    AM793 & $ 0.71\pm0.07 $   \\
12/01/2005 &  53382 &   A &    AM793 & $ 1.05\pm0.10 $   \\
30/01/2005 &  53400 &  AB &    AM793 & $ 0.99\pm0.10 $   \\
03/02/2005 &  53404 &  AB &    AM793 & $ 0.76\pm0.05 $   \\
11/02/2005 &  53412 &  AB &    AM793 & $ 1.36\pm0.23 $   \\
24/03/2005 &  53453 &   B &    AM793 & $ 1.06\pm0.23 $   \\
16/04/2005 &  53476 &   B &    AM793 & $ 0.81\pm0.05 $   \\
20/05/2005 &  53510 &   B &    AM793 & $ 0.82\pm0.06 $   \\
06/06/2005 &  53527 &   B &    AM793 & $ 0.94\pm0.08 $   \\
04/07/2005 &  53555 &  BC &    AM831 & $ 0.81\pm0.06 $   \\
22/08/2005 &  53604 &   C &    AM831 & $ 0.85\pm0.05 $   \\
02/09/2005 &  53615 &   C &    AM831 & $ 0.99\pm0.06 $   \\
16/10/2005 &  53659 &   C &    AM831 & $ 0.79\pm0.09 $   \\
24/12/2005 &  53728 &   D &    AM831 & $ 0.67\pm0.04 $   \\
11/02/2006 &  53777 &   A &    AM831 & $ 0.72\pm0.07 $   \\
22/04/2006 &  53847 &   A &    AM831 & $ 0.79\pm0.06 $   \\
30/07/2006 &  53946 &   B &    AM862 & $ 0.80\pm0.10 $   \\
25/11/2006 &  54064 &   C &    AM862 & $ 0.92\pm0.07 $   \\
05/03/2007 &  54164 &   D &    AM862 & $ 1.17\pm0.09 $   \\
03/10/2007 &  54376 &  AB &    AM901 & $ 0.86\pm0.07 $   \\
09/12/2007 &  54443 &   B &    AM901 & $ 1.22\pm0.18 $   \\
26/03/2008 &  54551 &   C &    AM901 & $ 1.45\pm0.14 $   \\
12/07/2008 &  54659 &   D &    AM952 & $ 0.92\pm0.10 $   \\
21/12/2008 &  54821 &   A &    AM952 & $ 0.94\pm0.08 $   \\
21/08/2009 &  55064 &   C &    AM952 & $ 1.17\pm0.11 $   \\
17/10/2009 &  55121 &   D &    AM983 & $ 1.06\pm0.04 $   \\
21/04/2010 &  55064 &   D &    AM983 & $ 1.24\pm0.05 $   \\
23/10/2010 &  55492 &   C &  10B-100 & $ 1.25\pm0.06 $   \\
22/03/2011 &  55642 &   B &  10B-100 & $ 1.22\pm0.03 $   \\
03/08/2011 &  55776 &   A &  10B-100 & $ 1.15\pm0.05 $   \\
02/01/2012 &  55928 &   D &  11B-001 & $ 1.25\pm0.05 $   \\
04/05/2012 &  56051 &   C &  11B-001 & $ 1.19\pm0.03 $   \\
23/07/2012 &  56131 &   B &  11B-001 & $ 0.96\pm0.04 $   \\
12/01/2013 &  56304 & A$\Rightarrow$D & 12B-005 & $ 0.95\pm0.03 $   \\   
23/07/2013 &  56496 &   C  & 12B-005 & $ 0.88\pm0.05 $   \\

\bottomrule
%\tabnotetext{a}{\mbox{Dates in format dd/mm/yyyy}}
%\tabnotetext{b}{\mbox{Flux is in mJy.}}
%\tabnotetext{c}{\mbox{Peak is in mJy$\!$ beam$^{-1}$.}}
\end{longtable}
\end{center}

\begin{table}[!h]\centering
  \setlength{\tabnotewidth}{0.5\columnwidth}
  \tablecols{5}
  % Stretch the space between table columns 
  \setlength{\tabcolsep}{2.8\tabcolsep}
  \caption{Equatorial coordinates of WR 112} \label{tab2}
 \begin{tabular}{lrrrr}
    \toprule
    Epoch & \multicolumn{1}{c}{$\alpha$} & \multicolumn{1}{c}{$\Delta \alpha$} & \multicolumn{1}{c}{$\delta$} & \multicolumn{1}{c}{$\Delta \delta$}\\
    & \multicolumn{1}{c}{$18^{\rm h} 16^{\rm m}$} & \multicolumn{1}{c}{$\times 10^{-3}$} & \multicolumn{1}{c}{$-18\arcdeg 58\arcmin$} & \multicolumn{1}{c}{$\times 10^{-2}$}\\  
       \midrule
2000.80 & $33\fs4923$ & 0.5 & $42\farcs352$ & 1.3 \\
2000.98 & $33\fs4943$ & 1.6 & $42\farcs382$ & 3.7 \\
2002.35 & $33\fs4904$ & 0.5 & $42\farcs321$ & 1.2 \\
2003.51 & $33\fs4950$ & 0.7 & $42\farcs301$ & 1.8 \\
2003.72 & $33\fs4897$ & 2.2 & $42\farcs346$ & 2.0 \\
2003.79 & $33\fs5026$ & 2.0 & $42\farcs040$ & 5.0 \\
2004.73 & $33\fs4904$ & 0.6 & $42\farcs379$ & 1.6 \\
2005.03 & $33\fs4904$ & 1.1 & $42\farcs359$ & 1.9 \\ 
2006.12 & $33\fs4928$ & 0.7 & $42\farcs381$ & 1.6 \\
2006.31 & $33\fs4906$ & 0.4 & $42\farcs407$ & 1.1 \\ 
2007.76 & $33\fs4927$ & 1.8 & $42\farcs288$ & 1.6 \\ 
2008.97 & $33\fs4918$ & 0.5 & $42\farcs395$ & 1.4 \\ 
2011.59 & $33\fs4903$ & 0.2 & $42\farcs414$ & 0.7 \\ 
2012.62 & $33\fs4907$ & 1.2 & $42\farcs393$ & 2.8 \\ 
2013.03 & $33\fs4902$ & 0.2 & $42\farcs398$ & 0.5 \\
\bottomrule
  \end{tabular}
\end{table}

\section{Conclusions}

We have analyzed the multi--epoch radio emission of the Wolf-Rayet
star WR~112 at 8.4 GHz, from the VLA archive. Our main results
are:

1. WR~112 is a variable radio source. We attempted to detect
periodicity in scales of days to years, with no
success.

2. The large structure of the infrared pinwheel nebulae is not visible
at 8.4 GHz at levels of 50 $\mu$Jy.

3. The proper motions derived here suggest significant peculiar motions for WR~112.

% 4. WR~112 present changes in positions during the epochs,
% some of the changes are abrupt and need to be studied further
% in the future.

\end{document}